\documentclass[aps,prl,twocolumn,groupedaddress]{revtex4}
\usepackage{graphicx}

\begin{document}


\title{Out-of-plane magnetization reversal processes of (Ga,Mn)As \\ with two different hole concentrations}



\author{K. Hamaya}
\affiliation{%
Institute of Industrial Science, The University of Tokyo, 4-6-1 Komaba, Meguro-ku, Tokyo 153-8505, Japan.
}%
\author{T. Taniyama}
\affiliation{%
Materials and Structures Laboratory, Tokyo Institute of Technology,\\ 4259 Nagatsuta, Midori-ku, Yokohama 226-8503, Japan. 
}%
\author{Y. Yamazaki}
\affiliation{%
Department of Innovative and Engineered Materials, Tokyo Institute of
Technology,\\ 4259 Nagatsuta, Midori-ku, Yokohama 226-8502, Japan.
}%



\date{\today}
\begin{abstract}
We study magnetization reversal processes of in-plane magnetized (Ga,Mn)As epilayers with different hole concentrations in out-of-plane magnetic fields using magnetotransport measurements. A clear difference in the magnetization process is found in two separate samples with hole concentrations of 10$^{20}$ cm$^{-3}$ and 10$^{21}$ cm$^{-3}$ as the magnetization rotates from the out-of-plane saturation to the in-plane remanence. Magnetization switching process from the in-plane remanence to the out-of-plane direction, on the other hand, shows no hole concentration dependence, where the switching process occurs via domain wall propagation. We show that the balance of $\left\langle 100 \right\rangle$ cubic magnetocrystalline anisotropy and uniaxial $[110]$ anisotropy gives an understanding of the difference in the out-of-plane magnetization processes of (Ga,Mn)As epilayers.

\end{abstract}


\maketitle

\section{INTRODUCTION}
One of the major steps in manipulating magnetic bits of ultra-high density magnetic memories is the development of electrical spin injection technology which is currently being studied intensively.\cite{Grollier} To understand the physics of the magnetization reversal using this approach, Mn-doped III-V magnetic alloy semiconductor (III-V MAS) is a promising material since the ferromagnetic properties which are induced by exchange interaction between hole carriers and Mn spins can be controlled through the hole carrier concentration.\cite{Matsukura,Dietl2,Keavney} Recent experiment using a spin field-effect transistor built with a III-V MAS channel layer, in fact, has demonstrated that magnetization reversal is achieved by reducing hole carrier concentration in the channel layer.\cite{Chiba1} Optical spin injection using circularly polarized light was also used to rotate the magnetization of (Ga,Mn)As on GaAs(001) from in-plane to out-of-plane orientation.\cite{Oiwa} However, the mechanism of the magnetization rotation in these systems still remains open, presumably due to the lack of an understanding of the out-of-plane magnetization reversal process in (Ga,Mn)As. Therefore, a clear description of the out-of-plane magnetization reversal process of (Ga,Mn)As is crucial to achieve the manipulation of magnetic bits using spin injection technique.

In order to understand magnetization reversal processes and magnetic domain structures, a number of magnetotransport studies have been reported so far,\cite{Tang,Wang,Hamaya2,Matsukura2,Goen,Baxter} where a change in longitudinal resistance in an in-plane magnetic field is well understood using anisotropic magnetoresistace (AMR) and the characteristic magnetic anisotropy in (Ga,Mn)As.\cite{Wang,Hamaya2,Matsukura2,Goen,Baxter} This interpretation agrees well with direct observation of magnetic domain structures and magnetization reversal process,\cite{Welp} clearly showing that magnetotransport measurement is a suitable tool for studying the magnetization reversal process in III-V MAS even for out-of-plane field orientation we are studying.

In this article, we discuss the out-of-plane magnetization reversal process of in-plane magnetized (Ga,Mn)As epilayers with two different hole concentrations using magnetotransport measurements. The magnetotransport properties from the out-of-plane saturation to the in-plane remanence, which are understood in terms of the magnetization rotation, are substantially varied with the hole concentration. In contrast, the magnetization reversal mechanism from the in-plane remanence to the out-of-plane saturation is found to be due to 90$^{\circ}$ domain-wall (DW) propagation which shows no hole concentration dependence of the samples. The contrasting results are attributed mainly to the $\left\langle 100 \right\rangle$ cubic magnetocrystalline anisotropy based on the zinc-blende type crystal structure, shape anisotropy due to the sample geometry and strain induced anisotropy along [001].

\section{SAMPLES AND EXPERIMENTS}
In-plane magnetized Ga$_{0.956}$Mn$_{0.044}$As (sample A) and Ga$_{0.912}$Mn$_{0.088}$As epilayers (sample B) 
with a thickness of 105 $\pm$ 3 nm and 103 $\pm$ 1 nm were grown on semi-insulating GaAs (001) substrates 
using low temperature molecular beam epitaxy at 235$^{\circ}$C and 195$^{\circ}$C, respectively. 
Both samples were deposited on a 400-nm thick GaAs buffer layer grown at 590$^{\circ}$C. 
To increase hole carrier concentration, sample B were subjected to post-growth annealing in N$_{2}$ atmosphere for 60 min at 250$^{\circ}$C.\cite{Hayashi} The hole carrier concentrations of sample A and sample B were measured by an electrochemical capacitance-voltage method at room temperature: 4.0 $\times$ 10$^{20}$ cm$^{-3}$ for sample A and 
1.5 $\times$ 10$^{21}$ cm$^{-3}$ for sample B.\cite{Moriya} High resolution x-ray diffraction analyses were carried out to estimate the film thickness and lattice parameters. For both samples, the in-plane lattice parameters $a_{//}$ along $a$ and $b$ axes show the same value of 5.653 \AA\ as that of GaAs bulk. Out-of-plane lattice parameter $a_{\bot}$, on the other hand,  is larger than that of the substrate due to tetrahedral compressive in-plane strain.\cite{Shen} The samples were confirmed to be coherently grown and fully-strained with no detectable relaxation and no additional misfit dislocations. Since the $\gamma$ angle between $a$ and $b$ axes in the film plane was measured to be $\gamma =$ 90$^{\circ}$ for both samples, the structural symmetry between [110] and [1$\overline{1}$0] is maintained even for the layer geometry. Magnetic properties were measured using a superconducting quantum interference device (SQUID) 
magnetometer. The Curie temperatures $T_{c}$ of sample A and sample B were estimated to be 60 K and 130 K from the temperature-dependent 
magnetization, respectively. 
The epilayers were patterned into 100 $\times$ 500 $\mu$m$^{2}$ rectangular bars along the [100] crystal axis in the film plane. 
Details of the fabrication process were described elsewhere.\cite{Hamaya2} All magnetotransport measurements were performed using a standard four-point ac method in a Quantum Design Physical Properties Measurement System at 4 K. 
For magnetotransport measurements, the samples can be rotated {\it in-situ} about [010] axis so as to change the angle $\theta$ between magnetic field direction in the plane normal to the film surface and [100] crystallographic axis.
\begin{figure}
\includegraphics[width=8.5cm]{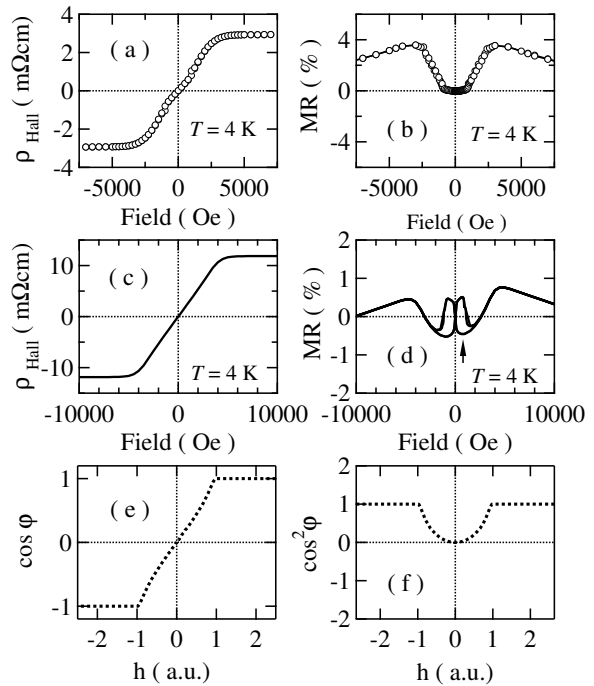}
\caption{ Field dependence of Hall resistivity and MR of sample A ((a), (b)) and sample B ((c, (d)) in out-of-plane magnetic fields at 4 K. 
Figures (e) and (f) show the calculation of $\cos \varphi$ and $\cos^{2}\varphi$ which corresponds to the Hall resistivity and MR, respectively.
}
\end{figure}

\section{RESULTS and DISCUSSION}
Magnetization vs. field curves ($M-H$ curve) of sample A and sample B were measured at field orientations [100], [110], and [1$\overline{1}$0] in the film plane (not shown here), clearly showing a typical cubic anisotropy along [100] at 4 K for sample A and a contrasting large uniaxial anisotropy along [110]\cite{Welp,Tang,Wang,Liu,Hamaya2,Hrabovsky,Kato,Sawicki2} for sample B. The marked [110] uniaxial anisotropy has been observed in (Ga,Mn)As with a hole concentration higher than $\sim$10$^{21}$ cm$^{-3}$.\cite{Kato,Sawicki2}
\begin{figure}[t]
\includegraphics[width=7.5cm]{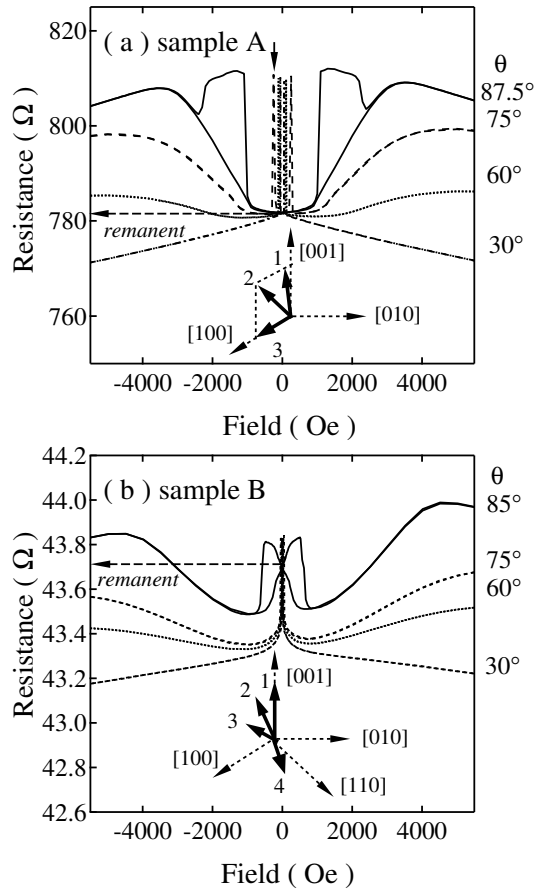}
\caption{Field-dependent MR of (a) sample A and (b) sample B at various applied field orientations $\theta_{}$ at 4 K. }
\end{figure}

Figure 1 shows the field-dependent Hall resistivity and magnetoresistance (MR) for sample A ((a), (b)) and sample B ((c), (d)). The MR (\%) is defined as ($R_{\rm H}-R_{0}$)/$R_{0}\times 100$, where $R_{\rm H}$ and $R_{0}$ are the resistance in a magnetic field of {\it H} and zero field, respectively. Magnetic fields are applied perpendicular to the film plane. In Fig. 1(a), it is clearly seen that the Hall resistivity in $H \leq$ $\pm$ 3500 Oe is associated with the magnetization process of (Ga,Mn)As. In general, the Hall resistivity is expressed as $\rho_{\rm Hall} =$ ($R_{0}/d$)$H$ $+$ ($R_{\rm S}/d$)$M$, where $R_{0}$ and $R_{\rm S}$ are the ordinary Hall coefficient and the anomalous Hall coefficient, respectively, $d$ is the film thickness, {\it H} is the applied field strength, and {\it M} is the magnetization. This shows that the Hall resistivity at a low field regime is attributed to anomalous Hall contribution while that at a high field regime originates from ordinary Hall contribution in the saturation magnetization state. The same explanation can also be applicable in Fig. 1(c) for sample B.

The MR curve in Fig. 1(b) can be explained by the anisotropic magnetoresistance (AMR).\cite{Esch,Matsukura} As shown in previous work,\cite{Hamaya2,Matsukura2,Wang,Baxter} the AMR of (Ga,Mn)As reflects on a change in the angle between the current direction [100] and the magnetization orientation, and shows the largest value when the magnetization is perpendicular to a current flow. This fact indicates that the MR curve in $H \leq$ $\pm$ 3500 Oe is associated with the magnetization process. The high-field MR ($H \geq$ $\pm$ 3500 Oe), on the other hand, shows a decrease with increasing magnetic fields. The decrease in the MR is likely due to a reduction in the scattering of hole carriers as seen in colossal magnetoresistance manganites.\cite{ODonnell} 

These features can be phenomenologically explained in terms of coherent rotation model as magnetic field is applied perpendicular to the film plane. If we assume strong in-plane anisotropy induced by compressive lattice strain\cite{Dietl2} and thin film structure, the total magnetostatic energy in perpendicular magnetic field can be expressed as 
$E=$ $K$$_{s}$sin$^{2}$ $\varphi$ $-$ $MH$cos ($\varphi$$-$$\theta_{}$), where
{\it K$_{s}$} is the in-plane anisotropy (shape-induced anisotropy + strain-induced anisotropy) constant and $\varphi$ is the angle between the magnetization direction and [100] direction. Using this expression, we can calculate the stable magnetization orientation by minimizing the total energy. Detailed calculation procedure is seen in previous reports.\cite{Hamaya3}
Figure 1(e) shows the calculated field dependence of $\cos\varphi$ which corresponds to the magnetization curve based on this model. Here $h$ is the magnetic field normalized with the
anisotropy field (2{\it K$_{s}$}/{\it M$_{}$}). As clearly seen, the curve in Fig. 1(e) is in good agreement with the experimental data shown in Figs. 1(a) and (c). The MR curve is also calculated as $\cos^{2}\varphi$ in Fig. 1(f).\cite{Hamaya3,Goen} Except for the high-field region, the MR curve in Fig. 1(b) is compatible with the calculation. The discrepancy between the experimental data and the calculation in the high field regime is again associated with a reduction in the scattering of hole carriers as described above. From these results, the magnetization process from saturation to the remanence is principally based on the magnetization rotation originating from the in-plane anisotropy for sample A.

Although very good agreement between the experimental data and calculation is obtained for sample A, a significant behavior can also be observed in the low field regime for sample B as seen in Fig. 1(d), where two clear peaks near $\pm$ 900 Oe are observed. To understand the low-field MR behavior, we show the MR curves of both samples at various field orientations $\theta_{}$ (0$^{\circ}$ $\leq$ $\theta_{}$ $<$ 90$^{\circ}$). Figure 2(a) shows the MR curves of sample A as a function of $\theta_{}$. Unlike the MR curve for $\theta_{} =$ 90$^{\circ}$ in Fig. 1(b), the MR curves exhibit hysteretic changes in $H \leq$ $\pm$ 3500 Oe while the resistance at remanence shows the same value of $\sim 782\ \Omega$ for all the field orientations. The remanent resistance, which is irrespective of the field orientation, clearly shows that the remanent magnetization is parallel to the current flow ([100] or [$\overline{1}$00]) in the film plane. When the field orientation tilts from the direction perpendicular to the film plane, a magnetization process illustrated in the inset of Fig. 2(a) likely occurs. A large contribution of $\left\langle 100 \right\rangle$ cubic magnetocrystalline anisotropy allows the magnetization process in (010) plane from 3500 Oe to 0 Oe and the magnetization aligns along [100] at remanence. After switching the magnetic field, the magnetization direction changes to [00$\overline{1}$], which is seen in the resistance jump to $\sim 810\ \Omega$, followed by a subsequent switching to [$\overline{1}$00] caused by [$\overline{1}$00] magnetic field component due to the tilted magnetic field direction. These switching processes are seen in the MR curve for $\theta_{} =$ 87.5$^{\circ}$ most clearly since a relatively large applied field is necessary to increase the in-plane component field sufficient to switch the magnetization to [$\overline{1}$00].
\begin{figure}[t]
\includegraphics[width=8.5cm]{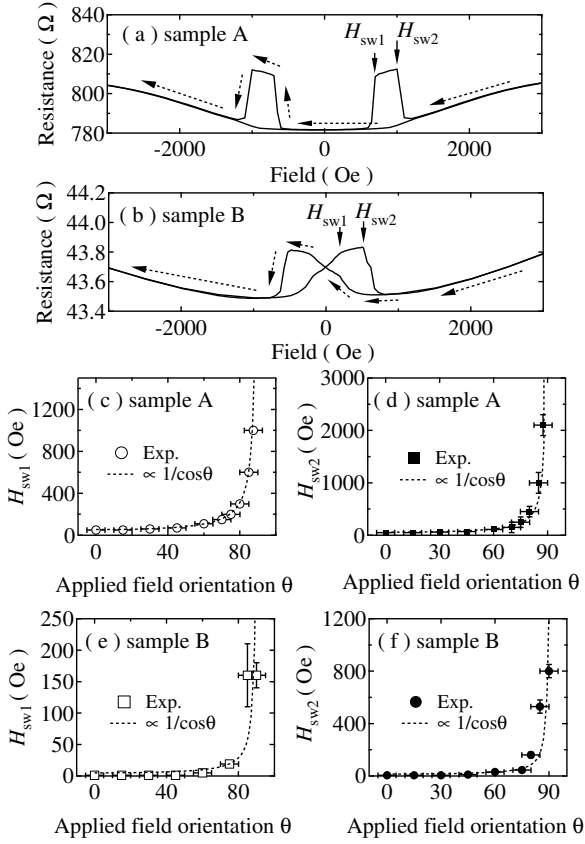}
\caption{MR curves of (a) sample A and (b) sample B measured at $\theta_{} =$ 85$^{\circ}$. {\it H}$_{sw1}$ and {\it H}$_{sw2}$ are switching fields as indicated by the arrows. {\it H}$_{sw1}$, {\it H}$_{sw2}$ $vs.$ $\theta_{}$ for sample A ((c),(d)) and sample B ((e),(f)). Dashed curves give a 1/cos$\theta_{}$ fit to the experimental data. }
\end{figure}

For sample B, on the other hand, the MR show a minimum value at around $\pm$500 Oe rather than at remanence as shown in Fig. 1(d). The distinct MR behavior can be explained in terms of a magnetization process based on the hole induced uniaxial anisotropy of sample B (hole concentration$\sim$ 10$^{21}$ cm$^{-3}$ ): a large [110] uniaxial anisotropy results in a different magnetization rotation process as shown in the inset of Fig. 2(b). When magnetic field decreases from 4500 Oe to 0 Oe, the magnetization direction now tilts away from (010) plane and aligns along [110] in the film plane at remanence. Since the AMR principally provides a minimum value for the magnetization direction which makes the smallest angle with the current direction, the minimum resistance at around 500 Oe indicates that the angle between magnetization and current flow is the smallest at around 500 Oe. When the field decreases from 500 Oe to the remanence, the magnetization rotates to a direction between 
 [100] and [110] in the film plane due to the anisotropy field along [110] so that the MR increases again due to an increase in the angle between magnetization and current flow. We also note that the out-of-plane saturation field (4500 Oe) of sample B is larger than that of sample A. The difference in the saturation field between these two samples is attributed mainly to a large shape anisotropy due to a factor of 2 larger magnetization of sample B. Also, a larger compressive strain in sample B could be another cause of the enhancement in the saturation field of sample B.

We plot the switching fields $H_{sw1}$ and $H_{sw2}$ from in-plane remanence to out-of-plane saturation as a function of $\theta_{}$ in Figs. 3(c)$-$(f) to understand the mechanism of the magnetization reversal process, where the switching fields $H_{sw1}$ and $H_{sw2}$ are defined as indicated by arrows in Figs. 3(a) and 3(b), respectively. 
The data for both samples well follow a $1/\cos\theta$ dependence, clearly showing that the magnetization process can be explained by domain wall propagation:\cite{Shigeto} this means that an effective in-plane field component which possesses $\cos\theta$ dependence pushes domain walls along the long bar axis. Also we note that the description of domain wall propagation is compatible with the magnetotransport data of other samples with different hole concentrations, indicating that the feature is universal for in-plane magnetized (Ga,Mn)As.
\begin{figure}[t]
\includegraphics[width=8.5cm]{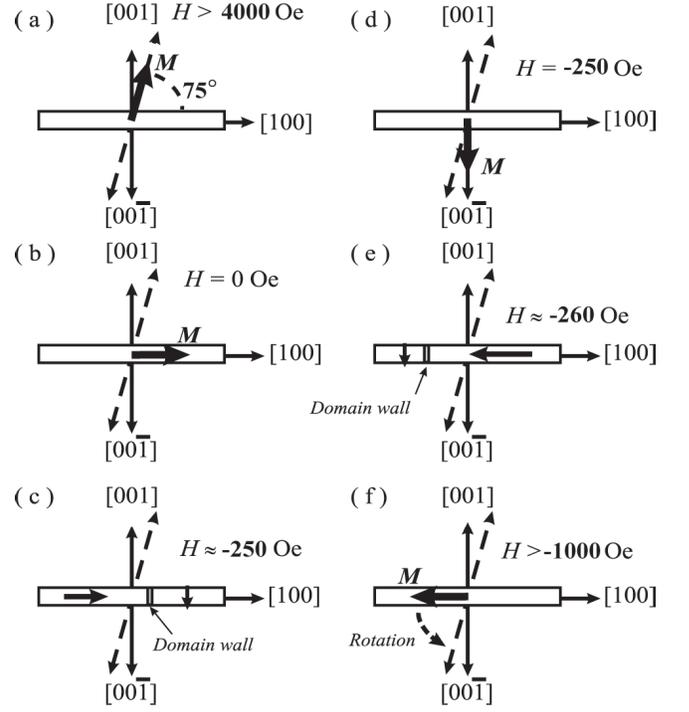}
\caption{Schematic illustration of magnetization switching process of sample A for $\theta_{} =$ 75$^{\circ}$. The nucleation and propagation of a 90$^{\circ}$ domain wall associated with $\left\langle 100 \right\rangle$ cubic magnetocrystalline anisotropy are shown. }
\end{figure}

From these results, we summarize the out-of-plane magnetization process of (Ga,Mn)As on the basis of AMR in Fig. 4, where the magnetization process of sample A at the field orientation $\theta_{} =$ 75$^{\circ}$ is illustrated. In $H\gtrsim 4000$ Oe (a), the magnetization saturates along the magnetic field. As the field decreases to $H = 0$ Oe (b), the magnetization lies in the film plane along [100]. With decreasing field down to -250 Oe (c), 90$^{\circ}$ DW nucleation occurs and propagates across the bar structure, resulting in the domain configuration perpendicular to the film plane (d). As the field decreases further (e), another 90$^{\circ}$ DW nucleates again and propagates immediately, hence a magnetic domain in the film plane becomes predominant. Finally, the magnetization rotates towards the field direction $\theta_{} =$ 75$^{\circ}$ with decreasing magnetic field (f). Similar magnetization process also occurs for the films with high hole concentrations except for the initial magnetization rotation process from the out-of-plane saturation to the in-plane remanence.

To explain the change in the magnetic anisotropy with increasing hole concentration, Sawicki {\it et al}. suggested the presence of trigonal distortion in the epilayer due to the anisotropic distribution of Mn ions.\cite{Sawicki2}  Also, the inhomogeneous distribution of Mn ions such as  Mn$_\mathrm{Ga}$-Mn$_\mathrm{Ga}$ and Mn$_\mathrm{Ga}$-Mn$_\mathrm{I}$-Mn$_\mathrm{Ga}$ is proposed in theoretical studies\cite{Mahadevan,Raebiger} and direct observation using scanning tunneling microscopy.\cite{Sullivan} Low-temperature growth of (Ga,Mn)As also suggests the presence of a cluster-like Mn distribution: recently, we have reported on a possible origin of complex in-plane magnetic anisotropy in (Ga,Mn)As on the basis of a cluster/matrix model,\cite{HamayaPRL,HamayaCond} where (Ga,Mn)As epilayers on GaAs(001) consist of ferromagnetic (Ga,Mn)As cluster-like regions with [110] uniaxial anisotropy and a ferromagnetic (Ga,Mn)As matrix with $\left\langle 100 \right\rangle$ cubic magnetocrystalline anisotropy. It was found that the Curie temperature of the clusters is higher than that of the matrix and the magnetic properties of the clusters dominate the magnetic anisotropy of (Ga,Mn)As with a high hole concentration. These results indicate that the hole concentration dependence of the magnetization rotation from the out-of-plane saturation to the in-plane remanence we observe originates mainly from the difference in the volume of these two phases at low temperatures. On the other hand, the magnetization reversal due to domain wall propagation is attributed to the intrinsic cubic $\left\langle 100 \right\rangle$ anisotropy as described above.

\section{CONCLUSIONS}
We have investigated the out-of-plane magnetization processes of (Ga,Mn)As epilayers with two different hole concentrations. A clear difference in the magnetization process is found in two samples with hole concentrations of 10$^{20}$ cm$^{-3}$ and 10$^{21}$ cm$^{-3}$ as the magnetization rotates from the out-of-plane saturation to the in-plane remanence. On the other hand, the magnetization switching from the in-plane remanence to the out-of-plane direction in a low-field regime is independent of hole concentration and we attribute the magnetization switching process to domain-wall (DW) propagation. Anisotropy field along [001] originating from the cubic magnetocrystalline anisotropy based on the zinc-blende type crystal structure of (Ga,Mn)As dominates the magnetization reversal process via DW propagation while shape anisotropy and strain induced magnetic anisotropy in the film plane also contribute to the magnetization rotation process from the out-of-saturation to the in-plane remanence.
\subsection{Acknowledgments}
Authors thank Prof. H. Munekata and Prof. Y. Kitamoto for kindly providing the opportunity to use his facilities to prepare the samples. 

\end{document}